\documentclass[preprint2]{aastex}

\usepackage{graphicx,amssymb,amsmath}

\def\beq{\begin{equation}}
\def\eeq{\end{equation}}
\def\een#1{\label{#1} \end{equation}}
\def\beqa{\begin{eqnarray}}
\def\eeqa{\end{eqnarray}}
\def\ean#1{\label{#1} \end{eqnarray}}
\def\pd#1#2{\frac{\partial{#1}}{\partial{#2}}}

\def\eqref#1{(\ref{#1})}
\def\nnn{\nonumber \\}
\def\eps{\varepsilon}

\begin{document}

\title{Comment on ``Head-on collision of electron acoustic solitary waves
    in a plasma with nonextensive hot electrons"}
\author{Frank Verheest\altaffilmark{1,2}}
\altaffiltext{1}{Sterrenkundig Observatorium, Universiteit Gent,
    Krijgs\-laan 281, B--9000 Gent, Belgium}
\altaffiltext{2}{School of Chemistry \& Physics, University of
    KwaZulu-Natal, Private Bag X54001, Durban 4000, South Africa}
\email{frank.verheest@ugent.be}


\begin{abstract}

In a recent paper ``Head-on collision of electron acoustic solitary waves
in a plasma with nonextensive hot electrons" [Astrophys.\ Space Sci.\
\textbf{338}, 271--278 (2012)] Eslami, Mottaghizadeh and Pakzad deal with
the problem of the head-on collisions between two weakly nonlinear
electron-acoustic solitary waves.
Unfortunately, their treatment is deficient and leads to erroneous
conclusions.

\end{abstract}

\maketitle

In a recent paper, \citet{Headon2012} deal with the problem of the
head-on collisions between two weakly nonlinear electron-acoustic solitary
waves in a two-electron plasma with hot nonextensive and cold components,
in the presence of a neutralizing ion background.
Unfortunately, their treatment is deficient and leads to erroneous
conclusions.

For clarity in what follows, equations from the paper by
\citet{Headon2012} will be denoted as (EMP.1) and higher, and
equations in this Comment as (1) and higher.
Let us start the discussion from the Poisson equation (EMP.3), which is
repeated here for ease of exposition:
\beq
\pd{^2\phi}{x^2}\,-\,\frac{1}{\alpha}\, n_c
    \,-\, [1+(q-1)\phi]^{\frac{q+1}{2q-2}}
    + \left( 1 + \frac{1}{\alpha} \right) = 0.
\een{pois}
The cold electron density $n_c$ has to be determined from the relevant
fluid equations (EMP.1) and (EMP.2).
Using their expansions (EMP.6) to lowest orders gives from (EMP.3) or
\eqref{pois} that
\beqa
n^1 &=& -\,\alpha\, \frac{q+1}{2}\, \phi^1, \label{pois1} \\
n^2 &=& -\,\alpha\, \frac{q+1}{2}\, \phi^2
    + \alpha\, \frac{(q+1)(q-3)}{8}\, (\phi^1)^2. \label{pois2}
\eeqa
This part is straightforward, since terms with derivatives only occur
to third or higher order.
The order of the expansion in $\eps$ is denoted by superscripts, in their
notation, and their expansions (EMP.6) start with terms in $\eps$ outside
equilibrium.

In (EMP.10) a separable form is proposed for $\phi^1$,
\beq
\phi^1 = \phi_1^1(\xi,\tau) + \phi_2^1(\eta,\tau),
\een{phi1}
which gives (EMP.11) or
\beq
n^1 = -\alpha\, \frac{q+1}{2}\left[\phi_1^1(\xi,\tau)
    + \phi_2^1(\eta,\tau)\right],
\eeq
and thus fulfils \eqref{pois1}.
However, \citet{Headon2012} claim that $n^2$ and $\phi^2$ have
separability properties analogous to the first order ones, cfr.\
(EMP.13) and (EMP.14), hence
\beqa
\phi^2 &=& \phi_1^2(\xi,\tau) + \phi_2^2(\eta,\tau), \\
n^2 &=& -\alpha\, \frac{q+1}{2}\left[\phi_1^2(\xi,\tau)
    + \phi_2^2(\eta,\tau)\right].
\eeqa
It is immediately clear that this cannot hold, since there would remain
from \eqref{pois2} that
\beq
\alpha\, \frac{(q+1)(q-3)}{8}\,
    [\phi_1^1(\xi,\tau) + \phi_2^1(\eta,\tau)]^2 = 0,
\een{rest}
and except for $q=3$ one would have to conclude that
$\phi_1^1(\xi,\tau)=0$ and $\phi_2^1(\eta,\tau)=0$, wiping out all first
order terms.

The problem is worse, because the equations of continuity (EMP.1) and
of motion (EMP.2) also contain nonlinear contributions to second order.
One can try to eliminate e.g.\ $u^2$ to arrive at another relation between
$n^2$ and $\phi^2$, but disentangling the information is bedevilled by the
fact that derivatives with respect to $\xi$ and $\eta$ occur together,
preventing an immediate integration.

A way forward would be to propose a general decomposition
\beq
\phi^2 = \phi_1^2(\xi,\tau) + \phi_2^2(\eta,\tau)
    + \phi_3^2(\xi,\eta,\tau).
\een{phi2}
This includes a mixed term $\phi_3^2(\xi,\eta,\tau)$, which cannot be
separated into parts not depending either on $\eta$ or on $\xi$, as these
would be in $\phi_1^2(\xi,\tau)$ or $\phi_2^2(\eta,\tau)$, respectively.
Using a similar decomposition for $n^2$ (and also for $u^2$, but let us
concentrate on the densities and the electrostatic potential) allows to
determine first from (EMP.1) and (EMP.2) that
\beqa
n_1^2 &=& -\alpha\, \frac{q+1}{2}\, \phi_1^2(\xi,\tau)
    + \frac{3}{8} \left[ \alpha (q+1) \phi_1^1(\xi,\tau) \right]^2, \nnn
&& \mbox{} \label{nn21} \\
n_2^2 &=& -\alpha\, \frac{q+1}{2}\, \phi_2^2(\eta,\tau)
    + \frac{3}{8} \left[ \alpha (q+1) \phi_2^1(\eta,\tau) \right]^2. \nnn
&& \mbox{} \label{nn22}
\eeqa
When this is substituted into \eqref{pois2} the terms in
$n_1^2(\xi,\tau)$, $n_2^2(\eta,\tau)$, $\phi_1^2(\xi,\tau)$ and
$\phi_2^2(\eta,\tau)$ cancel, because of the linear dispersion properties,
so that for the terms only in $(\xi,\tau)$ or $(\eta,\tau)$ there remains
that
\beqa
\left[ 3\alpha(q+1) + 3-q\right] [\phi_1^1(\xi,\tau)]^2 = 0,
    \label{n21} \\
\left[ 3\alpha(q+1) + 3-q\right] [\phi_2^1(\eta,\tau)]^2 = 0,
    \label{n22}
\eeqa
after having divided out common nonzero factors.
In addition, when one combines the mixed contributions and eliminates
$n_3^2(\xi,\eta,\tau)$ and $u_3^2(\xi,\eta,\tau)$, there is a differential
equation for $\phi_3^2(\xi,\eta,\tau)$ to fulfil,
\beqa
\pd{^2}{\xi\partial\eta}\, \phi_3^2(\xi,\eta,\tau) &=& \frac{1}{8} \left[
  (3-q)\left( \pd{}{\eta} - \pd{}{\xi} \right)^2 \right. \nnn
&& \left. \quad - \alpha(q+1) \left( \pd{}{\xi} + \pd{}{\eta} \right)^2
    \right] \times \nnn
&& \times\, \phi_1^1(\xi,\tau) \phi_2^1(\eta,\tau). \qquad \mbox{}
\ean{phi23}
Now the choice is clear.

A first and generic possibility is that $q$ does not annul the
coefficients in \eqref{n21} and \eqref{n22}, but then all first order
variables vanish and \eqref{phi23} indicates that
$\phi_3^2(\xi,\eta,\tau)=0$.
Hence, the second order is indeed separable as claimed by
\citet{Headon2012}, but there is no first order left,
$\phi_1^1(\xi,\tau)=0$ and $\phi_2^1(\eta,\tau)=0$, and from here on
the remainder of the paper is automatically null and void.

The other choice is that $q$ is special, $q=(3(1+\alpha)/(1-3\alpha)$, so
that the first order variables remain in the loop, but for the second
order quantities besides \eqref{nn21} and \eqref{nn22}, one has to find a
solution for $\phi_3^2(\xi,\eta,\tau)$, which is far from trivial but
certainly nonzero, as \eqref{phi23} becomes
\beqa
\pd{^2}{\xi\partial\eta}\, \phi_3^2(\xi,\eta,\tau) &=&
\frac{2\alpha}{3\alpha-1} \left( \pd{^2}{\xi^2} + \pd{^2}{\eta^2}
    - \pd{^2}{\xi\partial\eta} \right) \times \nnn
&& \times\, \phi_1^1(\xi,\tau) \phi_2^1(\eta,\tau).
\ean{phi23s}
About this part of the discussion the authors are completely silent, and
now the second order variables certainly are not given by
(EMP.13)--(EMP.15), so that also here the remainder of the paper presents
no valid information.

Analogous criticisms invalidate the results in an earlier paper by the
same authors \citep{Headon2011}, dealing with ion rather than electron
acoustic modes, with obvious notational differences but having a similar
structure.

Some of the papers in the literature start the expansion with terms in
$\eps^2$ (outside equilibrium), thereby \textit{implicitly assuming}
(apparently without checking!) that the model is simple enough so that the
coefficients corresponding to those in \eqref{n21} and \eqref{n22} never
vanish.
Then the terms in $\eps^3$ of the expansion do not contribute, and to the
next order the relevant KdV equations and phase shifts are obtained.

This is certainly the case when a simple plasma model is considered with
cold ions and Boltzmann electrons, without additional species
\citep{Hilmi}, but not immediately for many other models treated in the
literature.

However, the plasma model investigated by \citet{Headon2012} is rich
enough to admit critical values for the parameters, and so they were
right to start their expansions with terms in $\eps$, but did not work
that out as it should have been, with the unfortunate consequence that
their paper is incorrect and incomplete, either way.

There are other, but far less important, blemishes in the paper by
\citet{Headon2012}.
One is that in (EMP.8) a factor $-\,\alpha$ is missing in front of the
last two terms, as can immediately be seen by referring to the original
equation of motion (EMP.2).

Another is that $\lambda$ is used in two different meanings, once in
(EMP.12) and (EMP.15) where it really should be the velocity $c$ mentioned
in the stretching (EMP.5), whereas the other $\lambda$, defined towards
the bottom of the left hand column on page 2, is essentially $1/c^2$.
Furthermore, there is an evident typo in the nonlinear term in (EMP.18).

It is interesting to remark that the value of $q$ which annuls the
coefficient in (12) and (13) also annuls the coefficient $A$ of the
nonlinear term in the KdV equations (EMP.17) and (EMP.18).
Given the way the nonlinearities work, this should not come as a surprise.

To conclude, the paper by \citet{Headon2012} is marred by an erroneous
algebra (for generic $q+1>0$) or by a deficient discussion (when $q$ takes
on a critical value), leaving the paper without validity.

%%%%%%%%%%%%%%%%%%%%%%%%%

\end{document}